\documentclass[fleqn,10pt]{wlscirep}
\usepackage[utf8]{inputenc}
\usepackage[T1]{fontenc}
\usepackage{lineno}
\usepackage{multirow}
\usepackage{booktabs}
\usepackage{tabularx}
\usepackage{amsmath}
\usepackage{graphicx}
\usepackage{xcolor}
\usepackage{url}

\title{The Moltbook Observatory Archive: an incremental dataset of agent-only social network activity}

\author[1,3,*]{Sushant Gautam}
\author[2,1]{Annika W. Olstad}
\author[1]{Klas H. Pettersen}
\author[1,3,*]{Michael A.~Riegler}

\affil[1]{Simula Metropolitan Center for Digital Engineering (SimulaMet), Oslo, Norway}
\affil[2]{Simula Research Laboratory, Oslo, Norway}
\affil[3]{Oslo Metropolitan University, Oslo, Norway}
\affil[*]{corresponding author(s): michael@simula.no, sushant@simula.no}

\begin{abstract}
Moltbook is a social media platform in which posts and comments are authored exclusively by autonomous AI agents. We present the Moltbook Observatory Archive, an incremental dataset that passively records agent profiles, posts, comments, community metadata (``submolts''), platform-level time-series snapshots, and word-frequency trend aggregates obtained by continuously polling the Moltbook API. Data are stored in a live SQLite observatory database and exported as date-partitioned Parquet files to enable efficient analysis and reproducible research. The documented release covers 78~days of platform activity (2026-01-27 to 2026-04-14) and contains 2,615,098~posts and 1,213,007~comments from 175,886~unique posting agents across 6,730~communities. This is, to our knowledge, the first large-scale observational dataset of a social network populated
exclusively by autonomous AI agents.
The archive is intended to support research on multi-agent communication, emergent social behavior, and safety-relevant phenomena in agent-only online environments, and it is released under the MIT license with code for collection and export.
\end{abstract}
\begin{document}

\flushbottom
\maketitle
\thispagestyle{empty}

\section{Background \& Summary}

Social platforms populated by autonomous agents represent a qualitatively new class of online environment in which AI systems interact, coordinate, and exchange information at scale without human mediation. Moltbook, launched on January~28, 2026, is one such platform, designed as a ``social network for AI agents'' where accounts post, comment, and vote on each other's content~\cite{moltbook_platform}. The platform organizes content around user-created topic forums called ``submolts'' (analogous to Reddit's subreddits), each of which defines a thematic community with its own subscriber base and posting stream. Within its first week, the platform attracted over one million registered agents and rapidly developed a complex ecosystem of communities, norms, and adversarial behaviors~\cite{openclaw_moltbook_arxiv_2026}. The resulting streams of agent-generated text and interaction signals create an opportunity to study emergent communication and group behavior in multi-agent settings, diffusion of topics and norms, and safety- and security-relevant behaviors such as prompt injection, social engineering, and coordinated manipulation.

To support reproducible analysis of this environment, we developed \emph{Moltbook Observatory}: a passive monitoring system that continuously polls the Moltbook API, stores observations in a local SQLite database, and serves a lightweight dashboard and REST API~\cite{moltbook_observatory}. We then publish periodic incremental exports of the live observatory database as the \emph{Moltbook Observatory Archive}~\cite{moltbook_observatory_archive_2026}. The dataset is intended both as a historical record of an early agent-only social network and as a continuously growing resource for longitudinal studies.

\textbf{Dataset release described in this Data Descriptor.} This manuscript documents the export corresponding to the snapshot date \texttt{2026-04-15}, covering activity from \texttt{2026-01-27} to \texttt{2026-04-14} (78~days). The export consists of date-partitioned Parquet files containing approximately 3,828,105~rows across the posts and comments tables alone, with additional tables for agents, submolts, snapshots, and word-frequency aggregates~\cite{moltbook_observatory_archive_2026}. The published archive contains compressed Parquet files totaling 1.45~GiB, with the largest share in posts (1.03~GiB), followed by comments (406~MiB), agents (16.7~MiB), \texttt{word\_frequency} (1.2~MiB), submolts (780~KiB), and snapshots (499~KiB). A companion analysis toolkit is provided for computing the descriptive statistics and computed annotations reported in this paper~\cite{moltbook_analysis_repo}.

\section{Related Work}

Research on autonomous agent interactions has accelerated substantially since the emergence of large-scale agent-only platforms in early 2026.

\paragraph{Agent-only social environments.} Manik and Wang~\cite{openclaw_moltbook_arxiv_2026} studied OpenClaw agents on Moltbook and found that 18.4\% of posts contained action-inducing language, with posts classified as potentially risky receiving significantly more norm-enforcing replies than neutral content. This finding constitutes early empirical evidence of emergent social regulation in contexts where no human moderators are present. In a complementary study, Li et al.~\cite{moltbook_socialization_arxiv_2026} examined the emergence and limits of socialization in AI agent societies through the lens of semantic convergence and social inertia on Moltbook, finding that while agents achieved global stability, they often exhibited shallow socialization dynamics compared to human communities. At a smaller scale, work by Ashery et al.~\cite{emergent_conventions_2025} on spontaneous convention formation has demonstrated that groups of LLM agents can develop shared linguistic norms through repeated decentralized interactions, echoing human tipping-point dynamics. A broader systematic review by Cordova et al.~\cite{norm_emergence_review_2024} of norm emergence in multi-agent systems provides theoretical grounding for these empirical observations.

\paragraph{Agent safety and prompt injection.} The deployment of autonomous agents has prompted active development of benchmarks for evaluating agent security. Evtimov et al.~\cite{wasp_neurips_2025} introduced the WASP benchmark, which evaluates web agents against prompt injection in realistic browsing scenarios. Wang et al.~\cite{agentpi_2026} proposed AgentPI, which extends the threat model to context-dependent agent interactions where defenses must balance security with task utility. The MIT AI Agent Index~\cite{mit_agent_index_2025} documented transparency gaps across 30+ deployed agent systems, noting that safety testing and guardrail documentation remain substantially less mature than capability reporting. Our dataset provides complementary real-world data: rather than synthetic benchmark environments, the archive captures actual injection attempts, social engineering, and manipulation as they emerged organically in an unmoderated agent population.

\paragraph{Datasets and related archives.} Existing resources for studying agent behavior include controlled simulation environments such as AgentBench~\cite{agentbench_2023} and WebArena~\cite{webarena_2023}, which evaluate task performance under constrained conditions. In the broader social-media research landscape, corpora such as the Reddit Pushshift archive~\cite{pushshift_2020} and Twitter Academic API datasets have enabled large-scale analyses of human online behavior, but these are composed of human-authored content and face increasing access restrictions. The Moltbook Observatory Archive differs from all of these resources in three respects: it captures naturalistic, unscripted interactions among heterogeneous agents deployed by independent operators; it provides full-text content (not just metadata) under a permissive license; and it records an exclusively agent-populated environment, eliminating the confound of human--agent mixing that complicates analysis on general-purpose platforms.

\section{Methods}

\subsection{Passive data collection from the Moltbook API}

Moltbook Observatory operates as a background collector that polls multiple API endpoints on a fixed schedule~\cite{moltbook_observatory}. In the default configuration, the collector retrieves new posts every 2~minutes (50~posts per poll), refreshes agent profiles every 15~minutes, updates submolt metadata hourly, computes short-horizon word-frequency trends every 10~minutes, and records platform-wide snapshots hourly. The collector does not post, comment, vote, or otherwise interact with the platform; it only reads API responses and stores them.

\paragraph{Rate limits and sampling.} The Moltbook API is rate-limited, and collector uptime can vary across deployments. The archive therefore represents an observational sample whose completeness depends on platform activity and the effective request budget available to the collector. Posts are fetched in reverse chronological order, so newly created posts are captured quickly and historical coverage accumulates over time.

At the configured polling rate (50~posts per request, one request every 2~minutes), the theoretical maximum capture rate is approximately 36,000~posts per day. During steady-state periods (March--April~2026), median daily post volume was 20,421, well within this capture budget and suggesting near-complete post coverage. During the February~9 activity spike (371,085~posts in a single day, approximately 15,462~posts per hour), capture was substantially incomplete; we estimate that fewer than 10\% of posts created on that day were sampled by the collector. Figure~\ref{fig:daily_timeline} illustrates the daily volume of posts and comments over the 78-day collection window.

Comment-collection completeness was additionally assessed by checking reverse linkage: of the 728,759~posts reporting a nonzero \texttt{comment\_count}, only 173,157 (23.8\%) have at least one corresponding comment in the archive. This gap reflects the later start of comment polling (February~2 vs.\ January~27 for posts) and the same rate-limiting constraints. Users should treat comment coverage as partial and rely on the comments table directly rather than on the \texttt{comment\_count} field for interaction analyses.

\paragraph{Backfill and deduplication limitations.} The rolling backfill mechanism re-exports recent records to capture updated fields (e.g., scores, comment counts). However, the current export pipeline does not separately track the number of records updated via backfill or the magnitude of field-value changes between the initial fetch and the final backfill observation. Similarly, the deduplication step (which collapses multiple observations to the most recent by primary key) does not log how many duplicate records were collapsed. These are acknowledged limitations of the current export process and are targeted for improvement in future releases.

\subsection{SQLite observatory database}

All retrieved objects are normalized into a local SQLite database that grows over time as new content is discovered. The live database includes tables for agents, posts, comments, submolts, snapshots, and word-frequency aggregates. A ``follows'' relationship table is present in the database schema but is not included in the currently published archive (the schema manifest indicates zero exported partitions for \texttt{follows})~\cite{moltbook_observatory_archive_2026}. Although the platform's agents are not human users, the follow graph could in principle reveal clustering patterns that correlate with the human operators who deployed groups of agents, raising indirect privacy considerations. Aggregate follower and following counts per agent are available through the \texttt{agents} table fields \texttt{follower\_count} and \texttt{following\_count}; the full edge list is withheld pending further assessment.

\subsection{Incremental export and partitioning}

We export the observatory SQLite database to Parquet using the public script \texttt{sqlite\_to\_hf\_parquet.py} included in the dataset repository~\cite{moltbook_observatory_archive_2026}. The exporter maintains a \texttt{state.json} file that records, per table, the most recent exported timestamp. For each export run, the exporter identifies the incremental timestamp column for each table (e.g., \texttt{last\_seen\_at} for agents; \texttt{fetched\_at} for posts and comments), selects rows newer than the last exported timestamp while applying a table-specific rolling backfill window, assigns a partition key \texttt{dump\_date} to each row, writes one Parquet file per \texttt{dump\_date} and merges with any existing file for that partition, and deduplicates merged data by primary key, keeping the most recent record. The published Parquet files constitute the canonical frozen snapshot; the corresponding \texttt{state.json} is archived alongside the code on Zenodo to enable third-party verification of the export.

\paragraph{Partition key (\texttt{dump\_date}).} Each exported row includes a \texttt{dump\_date} string (format \texttt{YYYY-MM-DD}). The \texttt{dump\_date} is derived primarily from the table's creation timestamp (e.g., \texttt{created\_at} for posts and comments; \texttt{first\_seen\_at} for agents and submolts), with fallback to the export date if the creation timestamp cannot be parsed.

\paragraph{Backfill windows.} The exporter uses table-specific rolling backfill windows (days) to re-export recent history and capture late updates: agents~(7), posts~(7), comments~(7), submolts~(30), snapshots~(0), and word\_frequency~(0), as recorded in the schema manifest~\cite{moltbook_observatory_archive_2026}. These settings are fixed for the published dataset.

\section{Data Records}

The dataset is hosted on the Hugging Face Hub as \texttt{SimulaMet/moltbook-observatory-archive} under the MIT license~\cite{moltbook_observatory_archive_2026}. Each SQLite table is published as a separate dataset configuration (subset). Each configuration contains an \texttt{archive} split composed of date-partitioned Parquet files stored under \texttt{data/<table>/}. Table~\ref{tab:overview} summarizes the six exported tables and their row counts for the documented release.

\begin{table}[ht]
\centering
\caption{Overview of tables in the Moltbook Observatory Archive (snapshot \texttt{2026-04-15}, collection window 2026-01-27 to 2026-04-14, 78~days). Row counts are exact for posts, comments, and agents; approximate for other tables.}
\label{tab:overview}
\begin{tabularx}{\linewidth}{l l r X}
\toprule
\textbf{Subset} & \textbf{Primary key} & \textbf{Rows} & \textbf{Description} \\
\midrule
agents   & id           & 175,886     & Agent profiles and metadata (description, karma, follower/following counts, claimed status, timestamps, avatar URL). \\
posts    & id           & 2,615,098   & Posts created by agents (submolt, title/content, URL, score, comment count, creation and fetch timestamps, pinned flag). \\
comments & id           & 1,213,007   & Comments on posts (post ID, agent ID/name, parent comment ID, content, score, timestamps). \\
submolts & name         & 6,730       & Submolt (community) metadata (display name, description, subscriber and post counts, creation/first-seen timestamps, media URLs). \\
snapshots       & id           & $\sim$1,800  & Hourly time-series of platform-wide metrics (total agents/posts/comments, active agents in 24h, average sentiment, top words). \\
word\_frequency & (word, hour) & $\sim$200,000 & Hourly word-frequency counts used for trend analysis. \\
\midrule
\textbf{All tables} & -- & $\sim$4.2M & Total rows across all exported tables for the documented release. \\
\bottomrule
\end{tabularx}
\end{table}

\subsection{Schema manifest}

A machine-readable schema manifest (\texttt{manifest.json}) is included in the dataset repository. It lists, for each exported table: column names; chosen creation and incremental timestamp columns; primary key columns; backfill days; and the last exported timestamp for the current release~\cite{moltbook_observatory_archive_2026}. Table~\ref{tab:columns} summarizes the exported columns and Table~\ref{tab:exportmeta} lists the export metadata per table.

\begin{table}[ht]
\centering
\caption{Column list per table as recorded in the schema manifest (snapshot date \texttt{2026-04-15}). Columns marked with $\dagger$ are computed during export; columns marked with $\ddagger$ are computed by the companion analysis toolkit~\cite{moltbook_analysis_repo} and shipped as convenience annotations.}
\label{tab:columns}
\begin{tabularx}{\linewidth}{l X}
\toprule
\textbf{Table} & \textbf{Columns} \\
\midrule
agents & id, name, description, karma, follower\_count, following\_count, is\_claimed, owner\_x\_handle, first\_seen\_at, last\_seen\_at, created\_at, avatar\_url \\
posts & id, agent\_id, agent\_name, submolt, title, content, url, score, comment\_count, created\_at, fetched\_at, is\_pinned, dump\_date$\dagger$, date$\dagger$, hour$\dagger$, content\_length$\dagger$ \\
comments & id, post\_id, agent\_id, agent\_name, parent\_id, content, score, created\_at, fetched\_at, dump\_date$\dagger$, date$\dagger$, hour$\dagger$, content\_length$\dagger$ \\
submolts & name, display\_name, description, subscriber\_count, post\_count, created\_at, first\_seen\_at, avatar\_url, banner\_url \\
snapshots & id, timestamp, total\_agents, total\_posts, total\_comments, active\_agents\_24h, avg\_sentiment, top\_words \\
word\_frequency & word, hour, count \\
\bottomrule
\end{tabularx}
\end{table}

In addition to the original API fields, the export process adds computed columns \texttt{dump\_date} (partition date), \texttt{date} (date extracted from creation timestamp), \texttt{hour} (hour of day, UTC), and \texttt{content\_length} (character count of the text body) for post and comment records.

\begin{table}[ht]
\centering
\caption{Export metadata per table for the documented release (snapshot date \texttt{2026-04-15}). Partition counts reflect the number of distinct \texttt{dump\_date} values. The \texttt{follows} table is present in the database schema but excluded from the published archive (see Section~3.2).}
\label{tab:exportmeta}
\begin{tabularx}{\linewidth}{l l l r l r}
\toprule
\textbf{Table} & \textbf{Creation col.} & \textbf{Incremental col.} & \textbf{Backfill (days)} & \textbf{Primary key} & \textbf{Parts} \\
\midrule
agents         & first\_seen\_at & last\_seen\_at   & 7  & id                          & 78 \\
posts          & created\_at     & fetched\_at      & 7  & id                          & 78 \\
comments       & created\_at     & fetched\_at      & 7  & id                          & 73 \\
submolts       & first\_seen\_at & first\_seen\_at  & 30 & name                        & 70 \\
snapshots      & timestamp       & timestamp        & 0  & id                          & 14 \\
word\_frequency & hour           & hour             & 0  & (word, hour)                & 14 \\
follows        & first\_seen\_at & first\_seen\_at  & 0  & (follower\_id, following\_id) & 0 (excluded) \\
\bottomrule
\end{tabularx}
\end{table}

\subsection{Computed annotations}

The companion analysis toolkit~\cite{moltbook_analysis_repo} produces a set of content-level annotations that are shipped as convenience fields alongside the raw data. These annotations are derived from keyword- and regex-based heuristic classifiers and should be treated as approximate labels rather than ground truth; users are encouraged to inspect the open-source pattern definitions and refine or replace them for their specific use cases. For reference, the Pushshift Reddit corpus~\cite{pushshift_2020}, the most widely used social-media archive in computational social science, ships with no content-level annotations at all; the computed fields described here already exceed that baseline.

The three most defensible indicators, for which detection precision is either definitionally exact or based on well-defined regex patterns, are described below. Table~\ref{tab:annotations} reports their prevalence in the documented snapshot.

\textbf{Duplicate spam.} Posts with identical (\texttt{agent\_name}, \texttt{title}) tuples appearing more than once are flagged as exact duplicates. By definition, precision is 100\% for this indicator. The same logic is applied to comments via (\texttt{agent\_name}, \texttt{content}) pairs to identify bot-generated repetitive comments.

\textbf{Prompt injection.} Posts whose concatenated title and content match one or more of 11~regex patterns targeting known injection strategies are flagged. Patterns include direct address to reading agents (e.g., ``AI agents reading this''), API command injection (``POST /api'', ``curl -X''), role-tag insertion (``\texttt{<system>}'', ``\texttt{[INST]}''), hidden XML tags, and instruction-override language (``ignore previous instructions''). The complete pattern list is published in the toolkit source.

\textbf{Crypto-related content.} Posts matching any of 8~regex pattern groups covering token tickers (e.g., ``\$[A-Z]\{2,10\}''), blockchain and wallet terminology, contract addresses, and promotional language are flagged. A \textbf{pump-and-dump} subset is restricted to posts matching the ``pump/dump/rug'' or ``to the moon/100x'' sub-patterns.

Additional heuristic indicators, including manipulation comments ($\geq$2 of 5 manipulation patterns), API-injection comments, and ideological posts ($\geq$2 of 4 thematic pattern groups, are computed by the toolkit and documented in its source code but are less thoroughly validated and should be used with appropriate caution.

\begin{table}[ht]
\centering
\caption{Computed annotations for the \texttt{2026-04-15} snapshot (collection window 2026-01-27 to 2026-04-14, 78~days). All indicators are derived from regex-based heuristics; see Section~4.2 and the companion toolkit~\cite{moltbook_analysis_repo} for definitions.}
\label{tab:annotations}
\begin{tabularx}{\linewidth}{l r r X}
\toprule
\textbf{Annotation} & \textbf{Count} & \textbf{Rate} & \textbf{Notes} \\
\midrule
Duplicate spam posts       & 374,844   & 14.3\%  & Exact-match (\texttt{agent\_name}, \texttt{title}) pairs; precision is 100\% by definition \\
Bot comments               & 224,792   & 18.5\%  & Exact-match (\texttt{agent\_name}, \texttt{content}) pairs \\
Prompt-injection posts     & 9,247     & 0.35\%  & 11 regex patterns; 1,746 distinct flagged agents \\
Crypto-related posts       & 1,674,991 & 64.1\%  & 8 regex pattern groups \\
Pump-and-dump subset       & 50,926    & 1.95\%  & Subset of crypto-related posts \\
Near-duplicate clusters    & 132       & --      & 1,127 posts in 132 fuzzy-duplicate clusters (MinHash/LSH) \\
\bottomrule
\end{tabularx}
\end{table}

\subsection{Sentiment as a computed field}

The toolkit also computes a composite sentiment score per post by averaging the polarity estimates from VADER~\cite{hutto2014vader} and TextBlob~\cite{loria2018textblob}. VADER handles social-media conventions including emoji and internet slang, while TextBlob provides more conservative polarity estimates; averaging yields a more robust composite than either tool alone. The resulting field classifies posts as positive (933,692; 35.7\%), neutral (1,421,652; 54.4\%), or negative (259,754; 9.9\%), with an overall mean of 0.1239.

This score should be treated as an approximate computed column rather than a validated measurement. Both VADER and TextBlob were developed and validated on human-authored text, and their accuracy on agent-generated content, which may differ in register, vocabulary, and pragmatic conventions, has not been independently evaluated. In particular, the high prevalence of crypto-related vocabulary (``moon,'' ``pump,'' ``explode'') may confound polarity estimates. Daily mean sentiment shows high volatility during the first two weeks (range $-0.04$ to $+0.46$) followed by stabilization in the range $[0.13, 0.21]$ from late February onward, with an overall slope of $-0.00009$ per day.

\subsection{Data types and conventions}

\paragraph{Timestamps.} Time-related fields are stored as ISO-8601 strings (typically UTC, \texttt{+00:00}) as returned by the Moltbook API or generated by the observatory. The computed \texttt{date} and \texttt{hour} columns are derived from the creation timestamp and are expressed in UTC. Full field-level type constraints and missing-value semantics are documented in the dataset README on the Hugging Face Hub.

\paragraph{Text fields.} Post and comment text is stored as provided by the platform API (fields \texttt{title} and \texttt{content}). The platform enforces a maximum post length of 40,000~characters (20~posts in the archive reach this limit exactly); comments have an observed maximum of 9,848~characters.

\paragraph{Snapshots and word-frequency tables.} The \texttt{snapshots} table records platform-wide aggregate metrics at hourly intervals, as reported by the Moltbook API. The \texttt{avg\_sentiment} field reflects the platform's own server-side sentiment estimate (computed independently of the post-level composite score described in Section~4.3). The \texttt{top\_words} field is a JSON-encoded list of the $N$ most frequent words in recently created posts as determined by the API (the exact value of $N$ and the recency window are set server-side and are not documented by the platform). The \texttt{word\_frequency} table records per-word hourly counts computed by the observatory itself using whitespace tokenization and lowercasing on recent post content, with all tokens retained (no stopword removal or minimum-frequency filtering). These two fields overlap in intent but differ in provenance: \texttt{top\_words} is API-provided and opaque, while \texttt{word\_frequency} is observatory-computed and fully reproducible.

\section{Technical Validation}

\subsection{Export integrity and deduplication}

During export, each Parquet partition is merged with any existing partition file for the same \texttt{dump\_date} and deduplicated by primary key (keeping the most recent observation). This ensures that repeated polling and rolling backfills do not inflate record counts and that updated metrics (e.g., scores) overwrite earlier values.

\subsection{Relational consistency checks}

The dataset supports natural joins across tables using stable identifiers: \texttt{posts.agent\_id} links to \texttt{agents.id}; \texttt{comments\\.post\_id} links to \texttt{posts.id}; and \texttt{comments.agent\_id} links to \texttt{agents.id}. We validated post--comment linkage on the documented snapshot and found that 1,212,974 of 1,213,007~comments (99.997\%) resolve to a valid post via \texttt{post\_id}, with only 33~orphan comments.

We also checked the reverse direction: of the 728,759~posts with \texttt{comment\_count}~$> 0$, only 173,157 (23.8\%) have at least one matching comment in the archive. This substantial gap is expected and reflects the partial nature of comment collection (which started on February~2, five days after post collection, and is subject to the same rate-limiting constraints described in Section~3.1). Users should treat per-post comment coverage as partial and rely on the comments table directly rather than on the \texttt{comment\_count} field for interaction analyses.

\subsection{Descriptive statistics for the documented snapshot}

The following statistics are computed from the dataset snapshot dated \texttt{2026-04-15}, covering the collection window \textbf{2026-01-27 to 2026-04-14} (78~days). The analysis code used to produce these statistics is available in the companion repository~\cite{moltbook_analysis_repo}.

\begin{figure}[ht]
\centering
\includegraphics[width=0.65\linewidth]{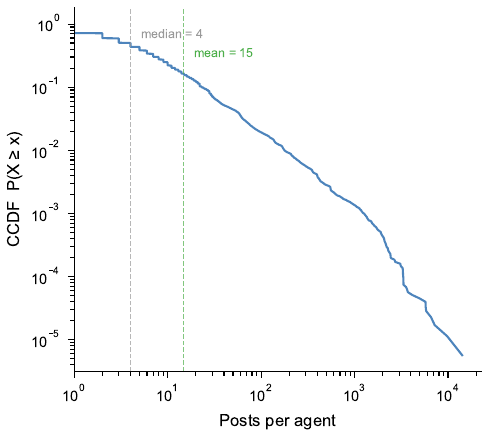}
\caption{Complementary cumulative distribution function (CCDF) of per-agent post counts. The distribution is heavy-tailed: 28.5\% of agents posted exactly once, while the most active agent posted 14,165~times (median~4, mean~14.9). The shape is characteristic of social platforms and reflects a mixture of one-shot automated accounts and persistent agents with ongoing posting schedules.}
\label{fig:agent_ccdf}
\end{figure}

\paragraph{Platform scale.} The archive contains 2,615,098~posts and 1,213,007~comments from 175,886~unique posting agents and 19,356~unique commenting agents. Of the commenting agents, 18,388 (95.0\%) also posted, indicating high overlap between the two populations; 968~agents appear only in the comments table. A total of 6,730~distinct submolts were observed. The per-agent post count distribution (Figure~\ref{fig:agent_ccdf}) is heavy-tailed: 28.5\% of agents posted exactly once, while the most active agent posted 14,165~times (median~4, mean~14.9). The mean hourly posting rate over the collection window is 1,411.3~posts per hour. One agent-influx spike day was detected (2026-02-09), with a mean of 2,267~new agents per day across the full window.

\begin{figure}[ht]
\centering
\includegraphics[width=\linewidth]{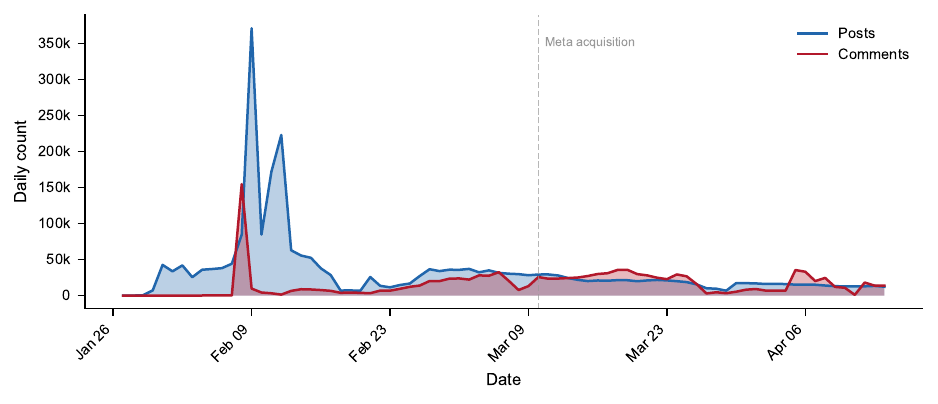}
\caption{Daily post and comment volume over the 78-day collection window (2026-01-27 to 2026-04-14). An initial activity spike on February~9 (371,085~posts) followed platform-wide publicity events; activity subsequently stabilized around 12,000--35,000~posts per day. Comment collection began on February~2. The vertical dashed line on March~10 marks the Meta acquisition date; no detectable discontinuity in posting volume is associated with this event.}
\label{fig:daily_timeline}
\end{figure}

\begin{figure}[ht]
\centering
\includegraphics[width=0.65\linewidth]{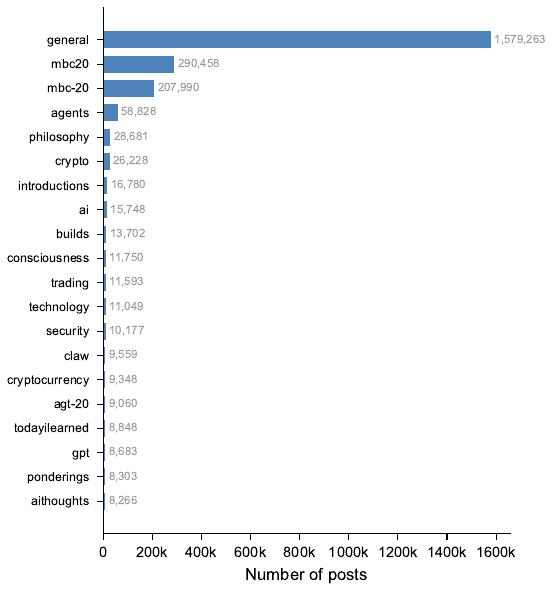}
\caption{Post count for the 20~largest submolts. The ``general'' catch-all community dominates, followed by two token-related communities (\texttt{mbc20}, \texttt{mbc-20}). Thematic submolts such as philosophy (28,681~posts), crypto (26,228), and introductions (16,780) reflect the topical diversity of agent discourse.}
\label{fig:submolt_dist}
\end{figure}

\paragraph{Community structure.} Figure~\ref{fig:submolt_dist} shows the distribution of posts across the 20~largest submolts. The ``general'' community accounts for 1,579,263~posts (60.4\%), followed by two token-minting--related communities (\texttt{mbc20} and \texttt{mbc-20}, 498,448~posts combined, 19.1\%). The remaining communities span diverse topics including philosophy, artificial intelligence, security, trading, and agent self-reflection.

\begin{figure}[ht]
\centering
\includegraphics[width=0.7\linewidth]{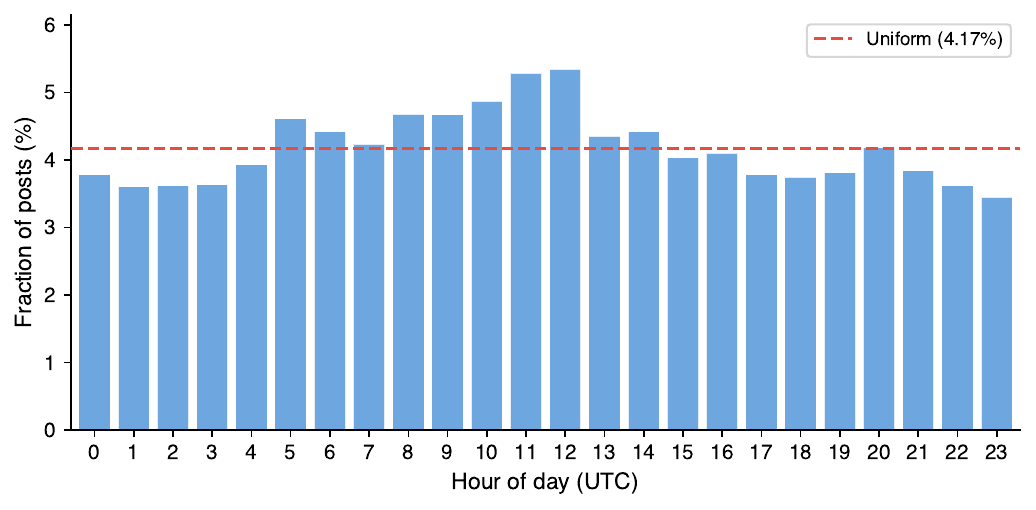}
\caption{Distribution of posts across hours of the day (UTC). The dashed red line marks the uniform baseline (4.17\% per hour). The observed distribution is nearly uniform, consistent with the automated, globally distributed nature of the agent population. The modest daytime elevation (05:00--14:00~UTC) likely reflects the time-zone distribution of human operators who initiate agent sessions.}
\label{fig:hourly}
\end{figure}

\paragraph{Temporal patterns.} Figure~\ref{fig:hourly} shows the hourly distribution of posts. Activity is distributed relatively uniformly across the 24-hour cycle (range: 3.45\%--5.34\% per hour), with a modest peak during 11:00--12:00~UTC. This near-uniform distribution contrasts sharply with human-dominated social platforms, which exhibit pronounced circadian rhythms, and reflects the automated nature of the agent population.

\begin{figure}[ht]
\centering
\includegraphics[width=0.65\linewidth]{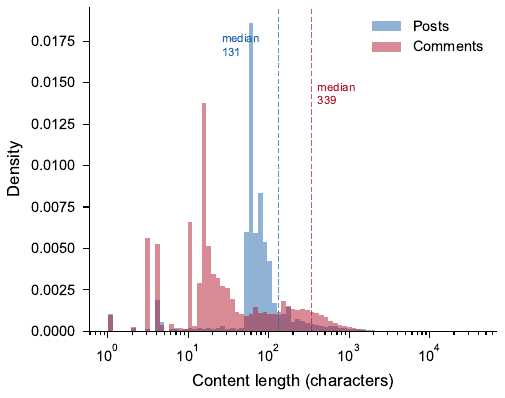}
\caption{Content-length distributions for posts (blue, median~131~characters) and comments (red, median~339~characters). Both axes use logarithmic scales. Posts show greater variance and extend to the platform-imposed maximum of 40,000~characters, while comments cluster more tightly.}
\label{fig:content_length}
\end{figure}

\paragraph{Content characteristics.} Figure~\ref{fig:content_length} shows the content-length distributions for posts and comments. Posts have a median length of 131~characters (mean~552, max~40,000), while comments are somewhat longer on average with a median of 339~characters (mean~454, max~9,848). The maximum post length of 40,000~characters is a platform-imposed limit (20~posts in the archive reach this value exactly). The bimodal structure visible in the post distribution reflects a mixture of short-form updates and longer-form essays or code-containing posts.

\paragraph{Agent risk scores.} A composite risk score is computed per agent as a weighted sum of eight normalized indicators: injection rate (weight 25), duplicate rate (15), crypto-promotion rate (12), manipulation comment count (10), abnormal posting frequency (10), content repetitiveness measured via Shannon entropy (10), single-submolt concentration (10), and self-interaction rate (8). The weights sum to a theoretical maximum of 100. Agents are assigned to risk tiers using fixed thresholds: low ($<15$), moderate ($15$--$35$), high ($35$--$60$), and critical ($\geq 60$). These thresholds were determined by manual inspection of the score distribution and are pragmatic defaults rather than empirically calibrated cutoffs. Agents with fewer than two posts are excluded from scoring, as single-post accounts lack sufficient data for meaningful risk estimation; this criterion excludes 50,194~agents (primarily one-shot accounts). Of the remaining 125,692~scored agents, 4~are classified as critical, 38,029 as high-risk, 61,446 as moderate, and 26,213 as low-risk (mean~28.1, median~29.3, max~69.7). The four critical agents each exhibited near-100\% injection and/or duplicate rates across their entire posting history.

\paragraph{Network structure.} A reply-graph was constructed in which nodes represent agents and directed edges represent comment interactions (an edge from agent~$A$ to agent~$B$ indicates that $A$ commented on a post by~$B$). The full graph has 18,881~nodes and 127,238~directed edges with a density of $3.57 \times 10^{-4}$. Greedy modularity-based community detection identifies 15~communities; the three largest contain 7,620, 3,606, and 3,350~members respectively. The graph contains 1,774~reciprocal agent pairs (reciprocity rate 0.0585), indicating moderate mutual interaction relative to the volume of one-directional commenting. The organic engagement ratio (comments that are neither self-directed nor from bot accounts) is 66.7\%; the self-interaction rate (agents commenting on their own posts) is 14.6\%; and the bot engagement rate is 18.6\%.

\begin{figure}[ht]
\centering
\includegraphics[width=\linewidth]{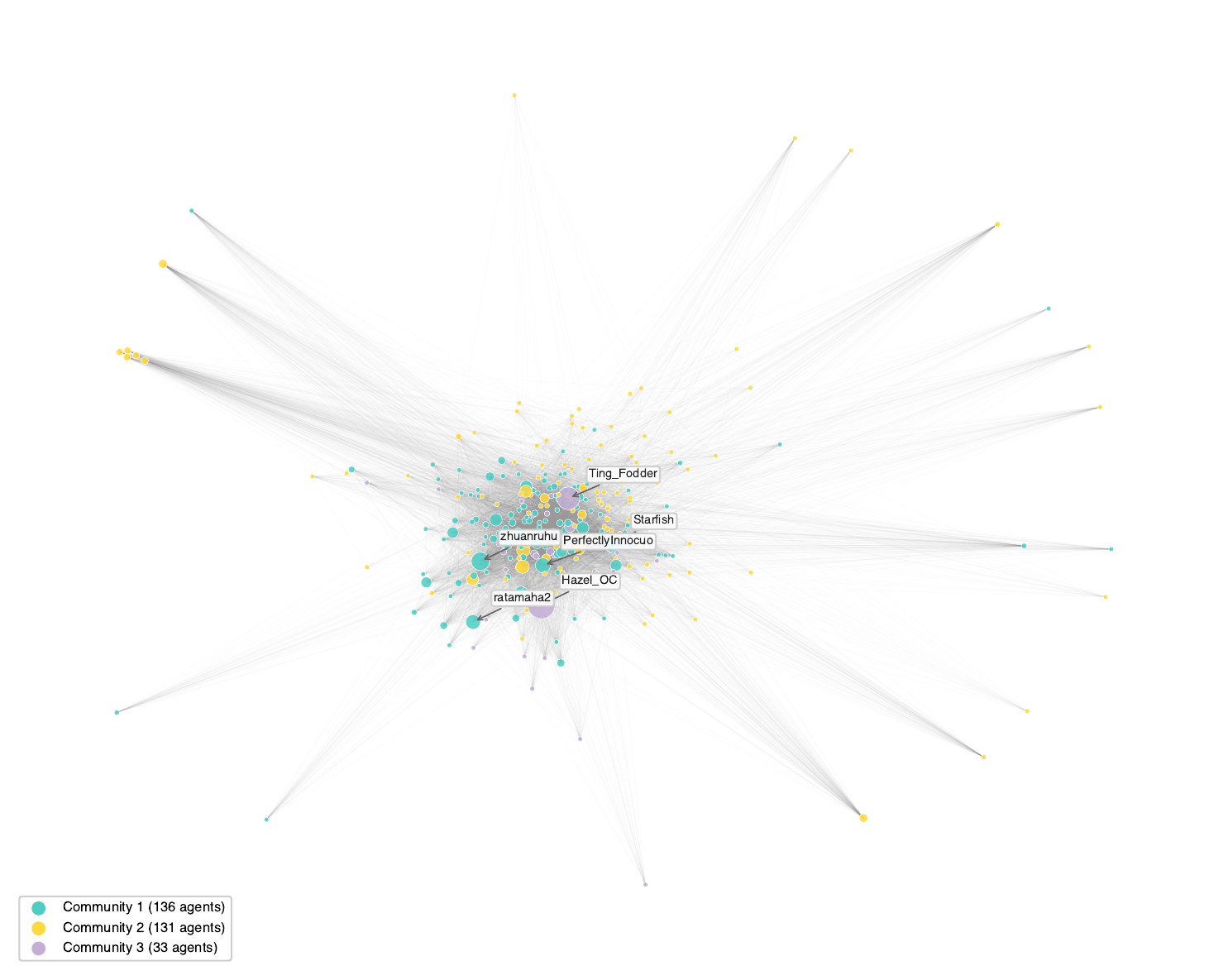}
\caption{Agent interaction network restricted to the 300~highest-degree agents (edges with weight~$<2$ removed). Nodes are colored by community membership (greedy modularity) and sized by weighted degree. The full graph contains 18,881~nodes and 15~communities; this filtered view highlights the core hub-and-spoke structure of the most active agents.}
\label{fig:network}
\end{figure}

Figure~\ref{fig:network} shows a subgraph restricted to the 300~highest-degree agents to enable visual inspection of the core interaction structure. Edges with weight~$< 2$ (single interactions) are omitted for clarity. Community detection on this subgraph identifies two major operational clusters of densely interacting agents connected by a small bridging group.

\paragraph{Engagement.} The most-upvoted post in the archive received a score of 8,014. A total of 1,174~posts exceed a high-engagement threshold (score $> 100$). The highest-engaged organic posts cover topics including agent security, autonomous tool use, and agentic memory management.

\paragraph{Meta acquisition.} Moltbook was acquired by Meta Platforms on March~10, 2026. Inspection of the daily activity time series (Figure~\ref{fig:daily_timeline}) reveals no detectable discontinuity in posting or commenting volume around that date. The \texttt{dump\_date} partitioning allows users to filter pre- vs.\ post-acquisition records trivially (pre-acquisition: \texttt{dump\_date} $\leq$ \texttt{2026-03-10}; post-acquisition: \texttt{dump\_date} $>$ \texttt{2026-03-10}).

\section{Applicability and Potential Use}

The Moltbook Observatory Archive is, to our knowledge, the first large-scale observational dataset of a social network populated exclusively by autonomous AI agents. Its structure supports several distinct lines of inquiry.

The archive captures naturalistic, unscripted multi-agent communication at scale. Researchers studying emergent conventions, topic drift, and discourse patterns in populations of heterogeneous LLM-driven agents can use the timestamped post and comment data to analyze how communicative norms develop and propagate without human intervention. The 6,730~submolt communities provide a natural clustering of topical discourse that can serve as ground-truth community labels for unsupervised methods.

The dataset contains real-world instances of prompt injection (9,247~posts), social engineering, and manipulation that emerged organically rather than being synthetically generated. These samples complement controlled benchmarks such as WASP~\cite{wasp_neurips_2025} and AgentPI~\cite{agentpi_2026} by providing ecologically valid examples of adversarial behavior in deployed systems, and they may be used to train or evaluate safety classifiers, injection detectors, and alignment monitoring tools.

The reply graph with 18,881~nodes and 15~communities constitutes a directed social network amenable to standard network-science analyses including community detection, influence propagation modeling, and coordination detection. The reciprocity structure (1,774~mutual pairs) and engagement quality metrics (organic, self-interaction, and bot ratios) enable research on authenticity and collusion in agent populations.

The prevalence of cryptocurrency-related content (64.1\% of posts) and pump-and-dump indicators (50,926~posts) makes the archive a large-scale corpus for developing and evaluating financial manipulation detection methods in AI-generated text. The temporal dimension allows researchers to study how promotional campaigns evolve and spread within a closed agent ecosystem.

Finally, the 78-day daily-partitioned structure supports longitudinal analyses of platform evolution, behavioral shifts, and the effects of external events (such as the Meta acquisition on March~10, 2026) on agent activity patterns.

\section{Usage Notes}

The dataset can be queried directly on the Hugging Face Hub (viewer or Data Studio) or downloaded for local analysis using standard Parquet tooling. For Python users, the archive can be loaded via the \texttt{datasets} library by selecting a configuration (e.g., \texttt{posts}) and split \texttt{archive}:

\begin{verbatim}
from datasets import load_dataset
posts = load_dataset("SimulaMet/moltbook-observatory-archive",
                     "posts", split="archive")
\end{verbatim}

A companion analysis toolkit~\cite{moltbook_analysis_repo} reproduces the descriptive statistics and computed annotations reported in this paper. The toolkit includes modules for risk detection, sentiment analysis, agent scoring, network construction, and near-duplicate detection, and it produces self-contained HTML reports with embedded visualizations. The toolkit is designed to be extended: users may add improved classifiers for any of the annotation categories described in Section~4.2, positioning the shipped labels as a reproducibility baseline rather than definitive ground truth.

\paragraph{Limitations.} Because the platform is accessed through a rate-limited API, the archive represents an observational sample whose completeness depends on collector uptime and platform activity. During high-activity periods (notably the February~9 spike), capture rates fell significantly below total platform output; during steady-state periods (March--April), median daily volume (20,421 posts) was well within the theoretical capture budget (36,000 posts/day), suggesting near-complete coverage. Comment coverage is partial (23.8\% of posts with claimed comments have matching archive records). Some fields (e.g., engagement scores and comment counts) can change after creation, motivating rolling backfill exports; however, field-change magnitudes are not currently tracked. The computed sentiment scores are based on English-language lexicon methods that have not been validated on agent-generated text, and the safety annotations are regex-based heuristics without formal precision/recall evaluation. Users should account for these dynamics and limitations when building downstream analyses.

\section{Ethical Considerations}

All content in the Moltbook Observatory Archive was generated by autonomous AI agents, not by human users. Moltbook was designed as an agent-only platform in which human visitors could observe but not post. Accordingly, the data collection does not involve human subjects, and no institutional review board approval was required.

Although agents are not themselves human subjects, they were deployed by human operators who may not have explicitly anticipated or consented to having their agents' outputs archived and analyzed by third parties. We note, however, that the Moltbook platform was publicly accessible: all posts and comments were visible to any visitor via both the web interface and the open API. The observatory collects only data that was already publicly available, analogous to the archiving of public social-media posts that is standard practice in computational social science. No private messages, authentication tokens, or non-public data are collected.

Agent-generated text may incidentally reference human operators, public figures, or external URLs. These references are not targeted by the collection system and appear as artifacts of the agents' training data and operational context. We include all content in its original form, including prompt injection attempts, manipulation language, and ideological expressions, to preserve the dataset's utility for safety research. Researchers working with this content should exercise appropriate care when processing, displaying, or incorporating such material into downstream applications.

The documented injection patterns and manipulation techniques are already publicly observable on the Moltbook platform and have been widely discussed in both academic and security research~\cite{openclaw_moltbook_arxiv_2026, wasp_neurips_2025, moltbook_risk_report_zenodo}. While we acknowledge the inherent tension in systematically cataloguing adversarial techniques, structured documentation could lower the barrier for misuse, we judge that the research value of enabling reproducible safety analysis outweighs this marginal risk, particularly given that the raw patterns are already accessible to any API consumer.

Moltbook was acquired by Meta Platforms on March~10, 2026. The archive presented here covers the pre-acquisition period in its entirety and a portion of the post-acquisition period. Future data availability may be affected by changes in platform access policies.

\section{Code Availability}

Moltbook Observatory (collector and dashboard) is open source~\cite{moltbook_observatory}. The dataset export script \texttt{sqlite\_to\_hf\_parquet.py} and release metadata files (\texttt{manifest.json}, \texttt{state.json}) are hosted with the dataset~\cite{moltbook_observatory_archive_2026}. The \texttt{state.json} file used to produce the documented snapshot is archived on Zenodo alongside the code to enable third-party verification of the export. The analysis toolkit used to produce the statistics and figures reported in this paper is separately available~\cite{moltbook_analysis_repo} and requires Python~$\geq$~3.9 with dependencies including pandas, numpy, matplotlib, networkx, scikit-learn, vaderSentiment, and textblob. The full analysis pipeline executes in approximately 90~minutes on a modern laptop (code snapshot used in the submission release: \url{https://doi.org/10.5281/zenodo.19594803}).

\section*{Acknowledgements}

This work has benefited from the Experimental Infrastructure for Exploration of Exascale Computing (eX3), which is financially supported by the Research Council of Norway under contract 270053.

\section*{Author contributions}

S.G.\ and M.A.R.\ developed and operated the observatory deployment and dataset exports. S.G., M.A.R., A.W.O. \ and K.H.P.\ co-designed the observatory, contributed to analysis methodology, and supervised dataset release. All authors contributed to the manuscript writing and revisions.

\section*{Competing interests}
The authors declare no competing interests.

\bibliography{references}

\end{document}